\documentclass[10pt,aps,twocolumn,superscriptaddress,longbibliography]{revtex4-1}

\usepackage{graphicx}  % Include figure files
\usepackage{amsmath}
\usepackage{amsmath,amssymb}
\usepackage{hyperref}
\usepackage{color}
\usepackage[sort&compress]{natbib}
\usepackage{amsmath,amssymb,MnSymbol,dsfont}

\def\PP{\mathbb{P}}
\def\RR{\mathbb{R}}

\begin{document}

\date{\today}
\title{Experimental Evidence of Hydrodynamic Instantons:\\
  The Universal Route to Rogue Waves} \author{Giovanni
  \surname{Dematteis}} \affiliation{Dipartimento di Scienze
  Matematiche, Politecnico di Torino, Corso Duca degli Abruzzi 24,
  I-10129 Torino, Italy} \affiliation{Dipartimento di Fisica,
  Universit\`a degli Studi di Torino, Via Pietro Giuria 1, 10125
  Torino, Italy} \author{Tobias \surname{Grafke}}
\affiliation{Mathematics Institute, University of Warwick, Coventry
  CV4 7AL, United Kingdom} \author{Miguel \surname{Onorato}}
\affiliation{Dipartimento di Fisica, Universit\`a degli Studi di
  Torino, Via Pietro Giuria 1, 10125 Torino, Italy} \affiliation{INFN,
  Sezione di Torino, Via Pietro Giuria 1, 10125 Torino, Italy}
\author{Eric \surname{Vanden-Eijnden}} \affiliation{Courant Institute,
  New York University, 251 Mercer Street, New York, NY 10012, USA}

\begin{abstract}
  A statistical theory of rogue waves is proposed and tested against
  experimental data collected in a long water tank where random waves
  with different degrees of nonlinearity are mechanically generated
  and free to propagate along the flume. Strong evidence is given that
  the rogue waves observed in the tank are \textit{hydrodynamic
    instantons}, that is, saddle point configurations of the action
  associated with the stochastic model of the wave system. As shown
  here, these hydrodynamic instantons are complex spatio-temporal wave
  field configurations which can be defined using the mathematical
  framework of Large Deviation Theory and calculated via tailored
  numerical methods. These results indicate that the instantons
  describe equally well rogue waves that originate from a simple
  linear superposition mechanism (in weakly nonlinear conditions) or
  from a nonlinear focusing one (in strongly nonlinear conditions),
  paving the way for the development of a unified explanation to rogue
  wave formation.
\end{abstract}

\maketitle

\section{Introduction}
\label{sec:intro}

A fascinating phenomenon observed in a wide class of nonlinear
dispersive systems is the occurrence of rogue waves with abnormally
large amplitude; they are found in sea surface gravity
waves~\cite{zakharov:1968,onorato-residori-bortolozzo-etal:2013},
nonlinear fiber optics~\cite{akhmediev2013recent},
plasmas~\cite{bailung2011observation} and Bose-Einstein
condensates. Rogue waves have received a lot of attention in the past
20 years, and different mechanisms for their formation have been put
forward, but a definite explanation has yet to be agreed
upon~\cite{kharif2008rogue,onorato2009statistical,%
  adcock2014physics,onorato-residori-bortolozzo-etal:2013,%
  fedele2016real,benetazzo2017shape}.  To settle this question,
studies in wave flumes or basins are interesting, because they permit
to create and measure wave states by means of mechanical wave
generators under controlled conditions meant to mimic (after
rescaling) those in the sea. The water surface in the tank can be
monitored accurately with high space-time resolution, and abundant
statistics can be collected. In one-dimensional experiments that mimic
an idealized long-crested rescaled sea, if the surface is sufficiently
energetic, nonlinear focusing effects take over linear dispersion and
are known to be responsible for increasing the likelihood of the rogue
waves. This leads to non-Gaussian fat-tailed statistics for their
amplitude~\cite{onorato:2001, onorato-residori-bortolozzo-etal:2013},
as opposed to the Gaussian statistics observed in the dispersive
regime.

In the present article, we propose a statistical theory of rogue waves
and test it against experiments performed in the one-dimensional
setting of the wave flume. We show that, in the full range of
experimental conditions tested, the rogue waves we observe closely
resemble {\it hydrodynamic instantons}~\cite{rajaraman1982solitons,
  dykman1994large, schafer1998instantons, falkovich1996instantons,
  grafke2015instanton, grafke2015relevance}: these are specific
spatio-temporal configurations of the wave field which we define
within the framework of large deviation theory (LDT) as the minimizers
of an action associated with the random wave model used to describe
the system; here we focus on the nonlinear Schr\"odinger equation
(NLSE) with random initial data but the approach is generalizable to
more complicated models. The finding that instantons explain
experimental rogue waves for a wide range of surface conditions in the
tank is striking because it offers a unified description of these
waves. In particular, our approach encompasses two of the main
existing theories for rogue wave creation: (i) the {\it theory of
  quasi-determinism}~\cite{lindgren1972local,boccotti2000wave} which
predicts that the rogue wave is created by linear superposition
effects and its shape is given by the autocorrelation function of the
wave field; (ii) the {\it semi-classical
  theory}~\cite{bertola2013universality,tikan2017universality} which
asserts instead that localized perturbations in the wave field can
lead to the formation of a Peregrine soliton via nonlinear focusing
instability. Our approach reconciles these two, apparently
incompatible, theories and smoothly interpolates between them as the
experimental control parameters are varied: when the nonlinear effects
are weak, the shape of the instantons converges to the autocorrelation
function predicted by the theory of quasi-determinism; and when the
nonlinear effects are strong, their shape converges to that of the
Peregrine soliton. Because the instanton calculus proposed in this
paper uses as limiting parameter the maximal wave amplitude itself,
without condition on model parameters or regimes in NLSE, it allows us
to assess the validity of the quasi-deterministic and semi-classical
theories by comparing them to the results of our approach in
appropriate regimes. Our approach could also be useful in the context
of other nonlinear theories for rogue waves based on NLSE, like
statistical approaches based on the Alber and the Wigner
equations~\cite{alber1978effects,
  onorato2003landau,stiassnie2008recurrent, ribal2013recurrent,
  gramstad2017modulational, athanassoulis2017localized}. We also
stress that the method proposed here can be generalized to the full
two-dimensional setting, as well as other relevant physical systems
where an understanding of extreme events is
important~\cite{aghakouchak2012extremes, coumou2012decade} but made
challenging by the complexity of the models involved combined with the
stochasticity of their evolution and the uncertainty of their
parameters~\cite{field2012managing,shepherd2016common,
  aghakouchak2012extremes, mohamad2016probabilistic,
  dematteis2019extreme}.  In this sense our approach adds to other
rare events methods~\cite{glasserman1999multilevel, juneja2006rare,
  cerou2007adaptive, giardina2011simulating, tailleur2007probing,
  VandenEijnden:2012ef, farazmand2017variational,
  ragone2017computation}.

The remainder of this paper is organized as follows: We introduce the
experimental setup in section~\ref{sec:experimental-setup}. In
section~\ref{sec:extr-event-filt}, we explain how we extract extreme
event data from the experimental measurements. Our approach based on
large deviation theory is presented in
Sec.~\ref{sec:theor-comp-inst}, where we also describe how we compute
the instanton for the rogue waves. Theory and experiment are then
compared in section~\ref{sec:valid-inst-descr}, with special focus on
the quasi-linear and highly nonlinear limiting cases. We conclude in
section~\ref{sec:conclusions} by discussing the implications of our
results in the context of a unified theory of rogue waves.
\begin{figure*}
  \centering \includegraphics[width=0.8\linewidth]{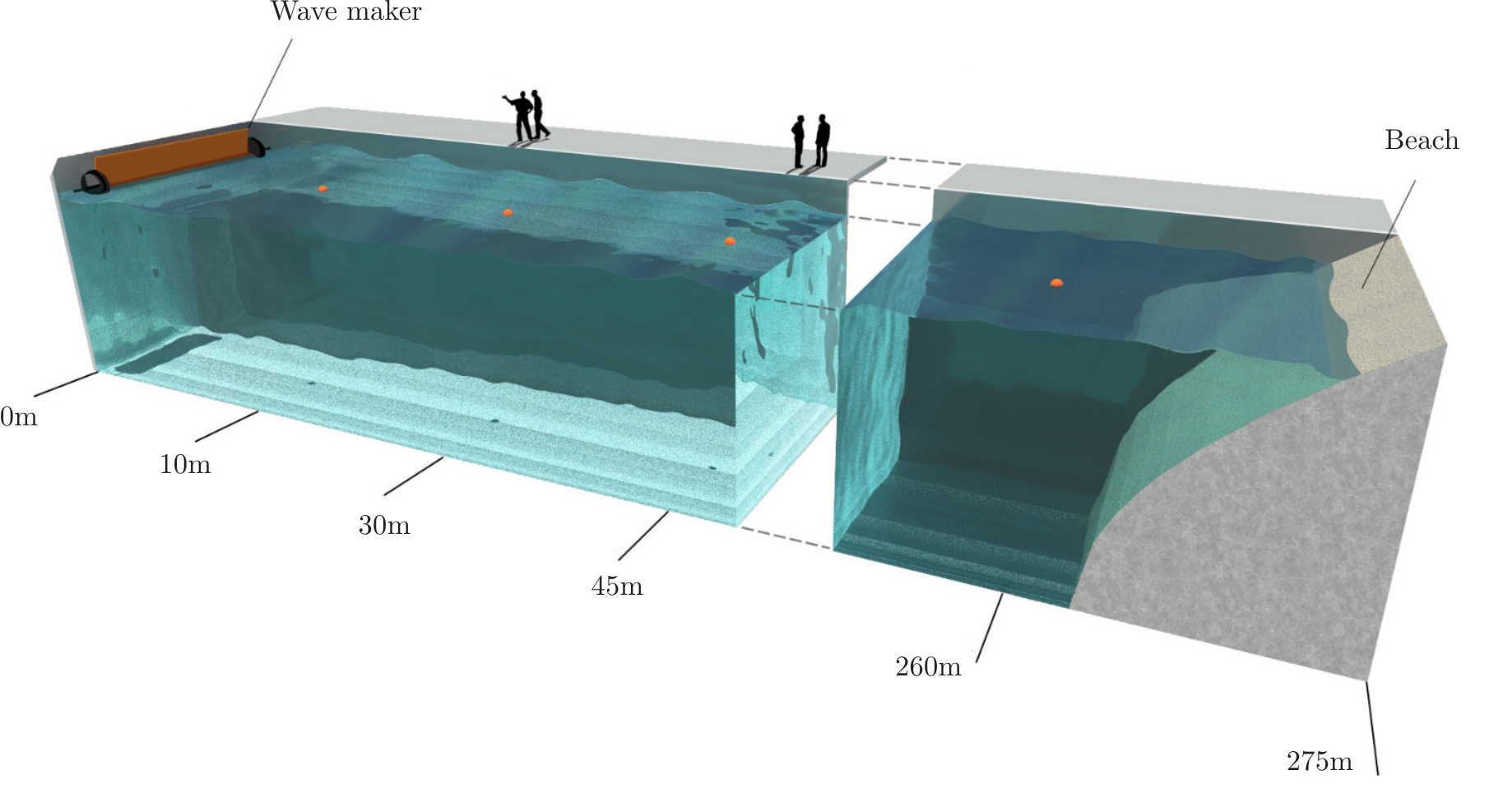}
  \caption{\textit{Wave flume experiment.} The wave maker generates a
    random wave field with stationary Gaussian statistics with the
    JONSWAP energy spectrum observed in the oceans. The planar wave
    fronts propagate along the water tank, where the surface elevation
    $\eta$ is measured by vertical probes.\label{fig:1}}
\end{figure*}

\section{Experimental setup}
\label{sec:experimental-setup}

The experimental data were recorded in the 270m long wave flume at
Marintek (Norway)~\cite{onorato2005modulational,onorato:2006},
schematically represented in Fig.~\ref{fig:1}. At one end of the tank
a plane-wave generator perturbs the water surface with a predefined
random signal. These perturbations create long-crested wave trains
that propagate along the tank toward the opposite end, where they
eventually break on a smooth beach that suppresses most of the
reflections. The water surface $\eta(x,t)$ is measured by probes
placed at different distances from the wave maker ($x$-coordinate).
The signal at the wave maker $\eta(x=0,t) \equiv \eta_0(t) $ is
prepared according to the stationary random-phase statistics with
{deterministic} spectral amplitudes $C(\omega_j)$:
\begin{figure}
  \centering \includegraphics[width=\columnwidth]{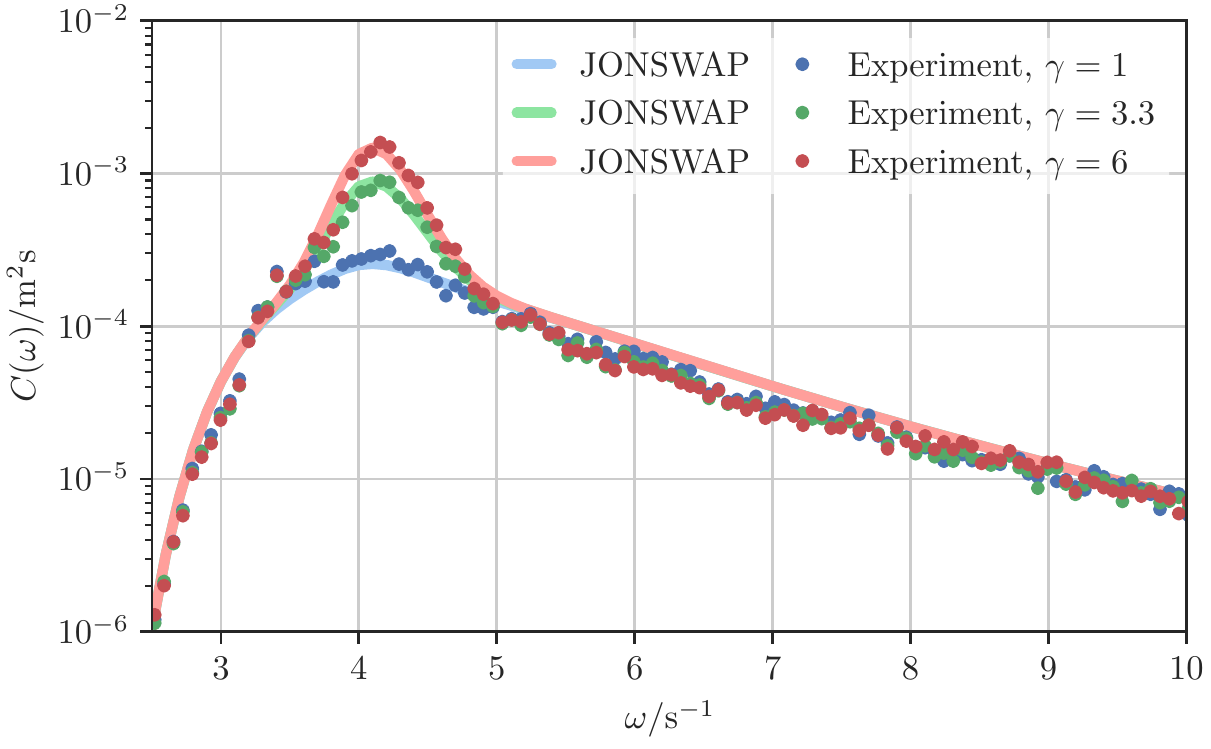}%
  \caption{JONSWAP spectra from Eq.~(\ref{eq:JONSWAP}) for the
    three experimental regimes of table~\ref{tab:1} (lines), compared
    to experimental measurements at the $x=10$m probe (dots).  These
    spectra remain roughly constant through the tank, except for small
    changes that are the signature of non-Gaussian effects that
    develop~\cite{onorato:2006}. \label{fig:jonswap-spectrum}}
\end{figure}
\begin{equation}\label{eq:1}
\eta^e_0(t)=\sum_{j=1}^N \sqrt{2
  C(\omega_j)\delta\omega}\cos(\omega_j t + \phi_j).
\end{equation}
Here the phases $\phi_j$'s are mutually independent random variables
uniformly distributed on $[0,2\pi]$,
$\delta\omega =\frac{2\pi}{\tau}$, $\omega_j=j\delta\omega$, and
$\tau$ is the time-series length. This guarantees that, for $N$ and
$\tau$ sufficiently large, $\eta^e_0(t)$ is approximately a stationary
Gaussian random field with energy spectrum $C(\omega)>0$, i.e.
\begin{equation}
  \label{eq:3}
  \begin{aligned}
    \langle \eta^e_0(t)\eta^e_0(t') \rangle &= \sum_{j=1}^N 
      C(\omega_j)\delta\omega\cos(\omega_j (t-t'))\\
    &\sim \int_0^\infty
    C(\omega) \cos(\omega(t-t')) d\omega,
  \end{aligned}
\end{equation}
where the bracket denotes expectation with respect to the random
phases $\phi_j$. In the experiment, $C(\omega)$ is taken to be the
JONSWAP spectrum~\cite{hasselmann:1973} of deep water waves observed
in the ocean,
\begin{equation}\label{eq:JONSWAP}
	C(\omega) = \frac{\alpha g^2}{\omega^5} \exp\left[
    -\frac54 \left( \frac{\omega_0}{\omega} \right)^4\right]
 \gamma^{
      \exp\left[
      -\frac{(\omega-\omega_0)^2}{2\sigma_J^2\omega_0^2}\right]}\,.
\end{equation}
Here $g=9.81$ms$^{-2}$ is the gravity acceleration, $\omega_0=4.19$
s$^{-1}$ is the carrier frequency (spectral peak), and
$\sigma_J=0.07$ if $\omega\le\omega_0$ and $\sigma_J=0.09$ if
$\omega>\omega_0$. These parameters are fixed for all sea states, and
we can use the dispersion relation of surface gravity waves in deep
water to obtain the carrier wave number $k_0=\omega_0^2/g = 1.79$
m$^{-1}$. The remaining parameters $\alpha$ and $\gamma$
in~\eqref{eq:JONSWAP} are dimensionless and vary according to weather
conditions. In the experiments, $\alpha=0.012$ throughout, while the
enhancement factor $\gamma$ ranges from 1 to 6, which is a realistic
range of values for the ocean measurements from calmer to rougher sea
states. In the water waves community, it is common to introduce the
significant wave height $H_s$, as a statistical measure of the average
wave height, here defined as
\begin{equation}
  \label{eq:17a}
H_s=4 \sigma=4 \left(\int_0^{\infty}C(\omega) d\omega\right)^{1/2},
\end{equation}
where $\sigma= \langle \eta_0^2\rangle^{1/2}$ is the standard
deviation of the surface elevation, which both depend on $\gamma$ as well
as the other parameters in~\eqref{eq:JONSWAP} that we keep fixed as
specified above. We also introduce a characteristic bandwidth~$\Omega$
of the JONSWAP spectrum defined as
\begin{equation}
  \label{eq:17b}
  \Omega = \text{width of $C(\omega)$ at half height.}
\end{equation}
Experimental data were collected for three different regimes: {\it
  quasi-linear} ($\gamma=1$, $H_s=0.11$~m), {\it intermediate}
($\gamma=3.3$, $H_s=0.13$~m), and {\it highly nonlinear} ($\gamma=6$,
$H_s=0.15$~m), see Table~\ref{tab:1}.  Note that these three regimes
have comparable significant wave heights~$H_s$, but the difference in
their enhancement factors~$\gamma$ has significant dynamical
consequences, as discussed in Sec. ~\ref{sec:theor-comp-inst} where we
introduce and explain the additional parameters $\epsilon$,
$L_{\text{lin}}$, and $L_{\text{Per}}$ listed in the
table. Experimental measurements of the spectrum for the three regimes
are depicted in Fig.~\ref{fig:jonswap-spectrum}.
 
For each set, we use data from $5$ time series, each of which is $25$
min long.  The surface elevation $\eta$ is measured simultaneously by
$19$ probes placed at different locations along the axes at the center
of the tank, recording data with a rate of $40$ measurements per
second. At each of two different positions ($x=75$~m and $x=160$~m)
two extra probes closer to the sides are used to check that the wave
fronts remain planar.

\begin{table}[h]
  \begin{center}
   \begin{tabular}{l||r|r|r|r|r|r}
	{\it Regime} & $\gamma$ & $H_s$(m) & $\Omega$(s$^{-1}$) &  $\epsilon$\ \  & $L_{\text{lin}}$(m) & $L_{\text{Per}}$(m)\\[1ex]
\hline {quasi-linear} & $1$ & $0.11$ & $2.12$ & $0.15$ & $8.9$ & $32$
    \\ {intermediate} & $3.3$ & $0.13$ & $0.90$ & $1.13$ & $46$ & $61$
    \\ {highly nonlinear} & $6$ & $0.15$ & $0.76$ & $2.23$ & $69$ & $65$\\
  \end{tabular}
  \caption{\small The relevant parameters in the three experimental
    regimes considered. The parameters $\gamma$, $H_s$, and $\Omega$
    are used to characterize the JONSWAP spectrum enforced by the wave
    maker. The parameter $\epsilon$ is used to quantify the strength
    of nonlinear versus dispersive effects in NLSE and is defined in
    Sec.~\ref{sec:model}. The two lengths $L_{\text{lin}}$ and
    $L_{\text{Per}}$ measure the typical scales over which these
    effects occur: they are defined in
    Section~\ref{sec:unif-pict-rogue} and are useful for the
    interpretation of Fig.~\ref{fig:4}}.\label{tab:1}
 \end{center}
\end{table}

\section{Extreme-event filtering: Extracting rogue waves
  from experimental data}
\label{sec:extr-event-filt}

\begin{figure*}
  \centering \includegraphics[width=1\linewidth]{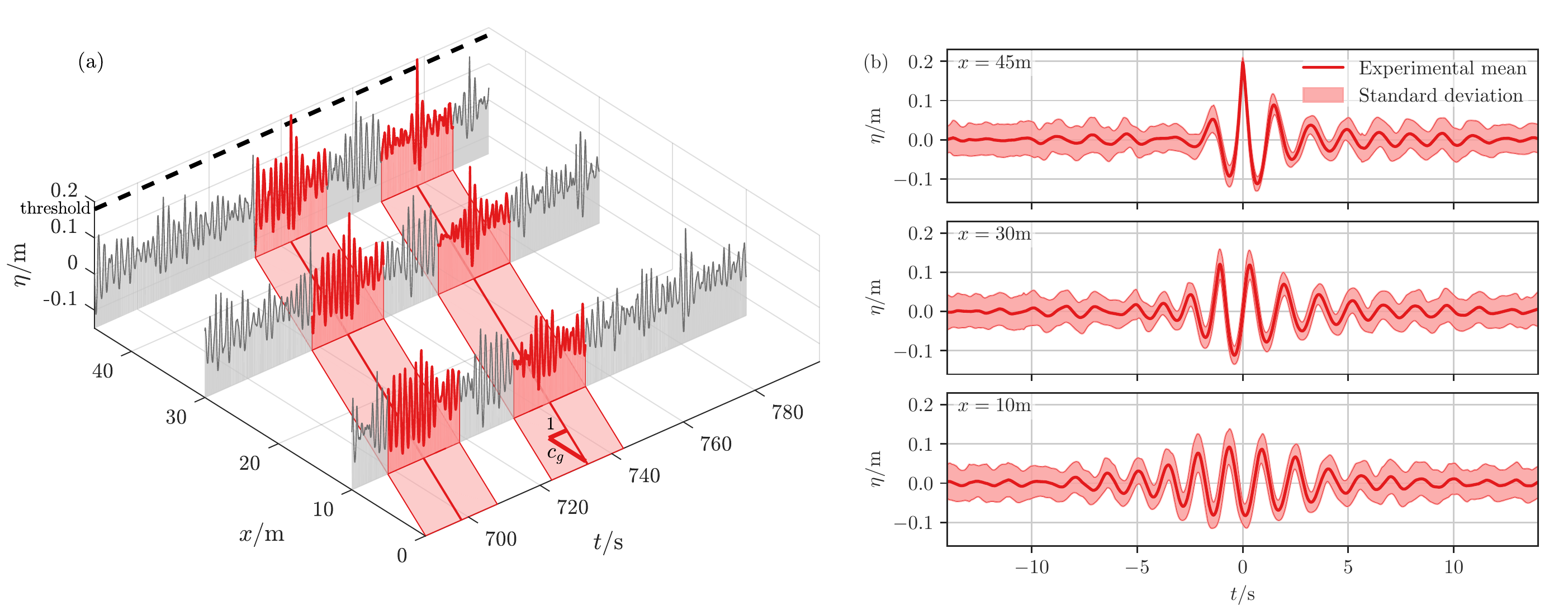}%
  \caption{(a) \textit{Extreme wave event selection.} At
    $x=45$~m, we monitor the temporal maximum of the experimental data
    series of $\eta$, record events reaching above a given threshold,
    and monitor the evolution of these events at probes located
    earlier in the channel. This is done within an observation time
    window centered at the maximum and following the wave packet with
    group velocity $c_g$; we repeat this for the whole time series to
    build a collection of extreme events and their evolution. (b)
    \textit{Mean extreme event.} The thick line shows the mean extreme
    event at different points along the channel, the shaded area a $1$
    standard deviation range around it. The {\it noise to signal}
    ratio is small in the focusing region, leading naturally to the
    question: Can we explain the common pathway by which these rogue
    waves are most likely to arise?\label{fig:filtering}}
\end{figure*}
To characterize the dynamics leading to extreme events of the water
surface, we adopt the following procedure: at a fixed location $x=L$
along the flume, we select small observation windows around all
temporal maxima of $\eta$ that exceed a threshold $z$. The choice of
the threshold $z$ is meant to select extreme events with a similar
probability for all sets: the values of $z=H_s=4\sigma$ for the
quasi-linear set, $z=1.1\,H_s=4.4\sigma$ for the intermediate set and
$z=1.2\,H_s=4.8\sigma$ for the highly-nonlinear set lead respectively
to $78$, $99$ and $88$ registered events where the maximum of the
surface elevation exceeds the threshold at the $45$ m probe,
$\eta(x=45$ m$,t)\ge z$.  We track the wave packet backward in space
and look at its shape at earlier points in the channel. This allows us
to build a collection of extreme events and monitor their
precursors. In Fig.~\ref{fig:filtering}a, we show two extreme events
at $x=45$~m obtained by this procedure, as well as their precursors at
$x=30$~m and $x=10$~m.  We analyze the statistical properties of these
extreme events by computing their average shape and the standard
deviation around it at the different positions along the channel,
obtaining the result shown in Fig.~\ref{fig:filtering}b for the
highly-nonlinear case.

\section{Theoretical description of rogue waves via instantons of
  NLSE}
\label{sec:theor-comp-inst}

We now explain how rogue waves can, within the framework of Large
Deviation, be described as instantons, that is, the minimizers of an
action functional associated with the nonlinear Schr\"odinger equation
with random initial data that we will use to describe the system's
evolution. In the linear case, as will be discussed later, this
minimization can be done analytically without much effort. When the
nonlinear effects matter, however, numerical computations are required
to perform the minimization.

\subsection{The  model}
\label{sec:model}

To avoid solving fully nonlinear water wave equations that are
complicated from both theoretical and computational viewpoints, it is
customary to use simplified models such as the Nonlinear Schr\"odinger
equation (NLSE). If we exclude very nonlinear initial data, it is
known that NLSE captures the statistical properties of one dimensional
wave propagation to a good degree of accuracy up to a certain
time~\cite{zakharov:1968, benjamin-feir:1967, onorato:2001,
  akhmediev2009extreme, onorato-residori-bortolozzo-etal:2013,
  chabchoub2011rogue} and it can be improved upon by using higher
order envelope equations \cite{trulsen-kliakhandler-dysthe-etal:2000,
  gramstad2011hamiltonian}. Because of their simplicity, NLSE and
extensions thereof have been successfully used to explain basics
mechanisms such as the modulational instability in water waves. With
the aim of capturing leading order effects, rather than describing the
full wave dynamics, here we restrict ourselves to the NLSE as a
prototype model for describing the nonlinear and dispersive waves in
the wave flume. Higher order models could in principle improve the
agreement between the theoretical instantons and the experimental
ones, but as demonstrated later, these corrections are negligible in
the wave flume experiment.

In the limit of \textit{deep-water}, \textit{small-steepness}, and
\textit{narrow-band} properties, the evolution of the system is
described, to leading order in nonlinearity and dispersion, by the
one-dimensional NLSE:
\begin{equation}\label{eq:NLSE}
  \frac{\partial \psi}{\partial x} + 2\frac{k_0}{\omega_0}
  \frac{\partial \psi}{\partial t} + {i} \frac{k_0}{\omega_0^2}
  \frac{\partial^2 \psi}{\partial t^2}+ 2 i {k_0^3} |\psi|^2\psi
  = 0\,.
\end{equation}
The NLSE describes the change of the complex envelope~$\psi\equiv\psi(x,t)$
that relates to the surface elevation via the Stokes series truncated
at second order:
\begin{equation}
  \label{eq:Stokes}
	 \eta = |\psi|\cos(\theta) + \tfrac12 k_0 |\psi|^2\cos(2\theta) +
     O(k_0^2 |\psi|^3)\,,
\end{equation}
where $\theta=k_0 x - \omega_0 t + \beta $ and $\beta$ is the phase of
$\psi$. In this expression the second order term can be neglected when
the field amplitude $|\psi|$ is small---this is the case near the wave
maker at $x=0$, where we will specify initial conditions for the
NLSE~\eqref{eq:NLSE}. However, this second order correction is
important when $|\psi|$ becomes large, i.e. when rogue waves develop.

The NLSE~\eqref{eq:NLSE} is written as an evolution equation in space
(rather than in time) in order to facilitate the comparison with
experimental data which are taken along the spatial extend of the
flume.  Consistent with the wave generator located at $x=0$, we
specify $\psi(x=0,t)=\psi_0(t)$ as initial condition
for~\eqref{eq:Stokes}, which we take to be a Gaussian random field
with a covariance whose Fourier transform is related to the JONSWAP
spectrum~\eqref{eq:JONSWAP}. Specifically, we set
\begin{equation}
  \label{eq:9}
  \psi_0(t) = \int_{-\infty}^\infty e^{i\omega t} \hat \psi_0(\omega) d\omega
\end{equation}
with 
$\hat \psi_0(\omega) $ Gaussian with mean zero and covariance
\begin{equation}
  \label{eq:10}
  \begin{aligned}
    \langle \hat \psi_0(\omega) \Bar{\hat\psi}_0(\omega')\rangle & = 
    C(\omega-\omega_0) \delta(\omega-\omega') \\
    \langle \hat \psi_0(\omega) \hat\psi_0(\omega')\rangle&= \langle
    \Bar{\hat \psi}_0(\omega) \Bar{\hat\psi}_0(\omega')\rangle = 0
  \end{aligned}
\end{equation}
where the bar denotes complex conjugation and $C(\omega)=C(-\omega)$ is the
JONSWAP spectrum defined in~\eqref{eq:JONSWAP}.  Since, to first order,
\begin{equation}
  \label{eq:11}
  \eta_0(t) = \tfrac12 \left(\psi_0(t) e^{-i\omega_0 t} + \bar \psi_0(t)
    e^{i\omega_0 t} \right) + O(k_0|\psi_0|^2)
\end{equation}
a direct calculation reported in Appendix~\ref{sec:app} shows that, to
that order, $\eta_0(t)$ is Gaussian with mean zero and covariance
$C(\omega)$. Note that in our setup the initial $\psi_0(t)$ is the
only source of randomness in the model. That is, we evolve $\psi_0(t)$
in space by the NLSE, and look for solutions $\psi(x,t)$ whose
elevation $\eta(x,t)$ exceed the threshold $z$ at spatial position
$x=L$, i.e.~satisfy $\eta(L,t)\ge z$ for some $t\ge0$ (using temporal
invariance we will later designate $t=0$ to be the point in time of
the extreme event).

The NLSE~\eqref{eq:NLSE} is Hamilton's equation
$i\left(\partial_{x}+(2k_0/\omega_0) \partial_t\right) \psi = \delta H
/ \delta\bar\psi$ associated with the Hamiltonian
$H=H_{\text{lin}}+H_{\text{nl}},$ with
\begin{equation}
  \label{eq:Ham}
  H_{\text{lin}} = -\frac{k_0}{\omega_0^2}\int_{-\infty}^\infty |\partial_t\psi|^2
  dt,\quad
  H_{\text{nl}}=k_0^3 \int_{-\infty}^\infty |\psi|^4 dt.
\end{equation}
In order to quantify the magnitude of the nonlinearity of the
wavefield, we use the ratio $\epsilon$ between the nonlinear energy
$H_{\text{nl}}$ and the free particle linear energy $H_{\text{lin}}$.
To this end, we use dimensional analysis to estimate
$|\partial_t\psi|^2=O(\Omega^2 H_s^2)$ and $|\psi|^4=O(H_s^4)$, where
averaged wave height~$H_s$ and the characteristic frequency~$\Omega$
are defined in~\eqref{eq:17a} and~\eqref{eq:17b}, respectively . This
gives
\begin{equation}
  \label{eq:eps}
  \epsilon = \frac{H_{\text{nl}}}{H_{\text{lin}}} =\left(\frac{\omega_0}{\Omega}
    k_0 H_s\right)^2.
\end{equation}
The values of $\epsilon$ obtained this way are given in
Table~\ref{tab:1} for the three regimes analyzed: quasi-linear,
intermediate, and highly nonlinear.  We stress that other definitions
of the nonlinearity parameter are possible, differing by a constant
factor---the important information is the relative magnitude
of~$\epsilon$ in the different regimes.  We also stress that the
values of~$\epsilon$ are used to interpret the results, but the
instanton calculations described next in
Sec.~\ref{sec:instanton-theory} are performed in the same way for all
values of~$\epsilon$.

\subsection{Large Deviation Theory and Instanton Calculus}
\label{sec:instanton-theory}

Our analytical and computational descriptions of rare events rely on
{\it instanton theory}. Developed originally in the context of quantum
chromodynamics \cite{schafer1998instantons}, at its core lies the
realization that the evolution of any stochastic system, be it quantum
and classical, reduces to a well-defined (semi-classical) limit in the
presence of a small parameter. Concretely, the simultaneous evaluation
of all possible realizations of the system subject to a given
constraint results in a (classical or path-) integral whose integrand
contains an action functional $S(\psi)$. The dominating realization
can then be obtained by approximating the integral by its {\it saddle
  point approximation}, using the solution to $\delta S(\psi^*)/\delta
\psi=0$. This critical point $\psi^*$ of the action functional is
called the {\it instanton}, and it yields the maximum likelihood
realization of the event. This conclusion can also be justified
mathematically within Large Deviation Theory.

Specifically, we are interested in the probability
\begin{equation}\label{eq:2}
	P_L(z) \equiv \PP(\eta(L,0)\ge z)
\end{equation}
i.e.~the probability of the surface elevation at position $L$ at an
arbitrary time $t=0$ exceeding a threshold $z$. This probability can
in principle be obtained by integrating the distribution of the
initial conditions over the set
\begin{equation}
\Lambda(z)=\{\psi_0: \eta(L,0))\ge z\},\label{eq:13}
\end{equation}
i.e.~the set of all initial conditions $\psi_0$ at the wave maker
$x=0$ that exceed the threshold $z$ further down the flume at
$x=L$. Since the initial field $\psi_0(t)$ is Gaussian, consistent
with \eqref{eq:10} the probability~\eqref{eq:2} can therefore be
formally written as the path integral
\begin{equation}
  \label{eq:pathintegral}
  P_L(z) = Z^{-1} \int_{\Lambda(z)} \exp(-\tfrac12 \|\psi_0\|^2_C)\,D[\psi_0]\,,
\end{equation}
where $Z$ is a  normalization constant and we defined
\begin{equation}
  \|\psi_0\|_C^2 = \int_{-\infty}^{\infty} \frac{|\hat
    \psi_0(\omega)|^2 }
  {C(\omega-\omega_0)} d\omega
\end{equation}
where
$\hat \psi_0(\omega) = 1/(2\pi) \int_{-\infty}^\infty \psi_0(t)
e^{-i\omega t} dt$ is the Fourier transform of $\psi_0(t)$. The
functional integral~(\ref{eq:pathintegral}) can be given a precise
mathematical meaning in several ways. For example, we can project the
initial field onto finitely many modes, in which
case~\eqref{eq:pathintegral} reduces to a regular integral over these
modes.  However, even if we were to perform this projection, the
integration is hard to perform in practice. This is because the set
$\Lambda(z)$ defined in~\eqref{eq:13} has a very complicated shape in
general, that depends non-trivially on the nonlinear dynamics
of~(\ref{eq:NLSE}) since it involves the field at $x=L>0$ down the
flume rather than $x=0$.  One way around this difficulty is to
estimate the integral~(\ref{eq:pathintegral}) via Laplace's
method. This strategy is the essence of {\it Large deviation theory}
(LDT), or, equivalently, \emph{instanton calculus}, and it is
justified for large~$z$, when the probability of the set $\Lambda(z)$
is dominated by a single $\psi_0$ contributing most to the integral
(see~\cite{dematteis2018rogue,dematteis2019extreme}). The optimal
condition leads to the constrained minimization problem
\begin{equation}
  \label{eq:4}
  \tfrac12\min_{\psi_0\in \Lambda(z)}\, \| \psi_0\|^2_C\equiv I_L(z)\,,
\end{equation}
and gives the large deviation estimate for Eq.~\eqref{eq:2},
\begin{equation}
  \label{eq:5}
  P_L(z) \asymp \exp\left(-I_L (z) \right)\,.
\end{equation}
where the symbol $\asymp$ means asymptotic logarithmic equivalence,
i.e.~the ratio of the logarithms of the two sides tends to 1 as
$z\to\infty$, or, in other words, the exponential portion of both
sides scales in the same way with~$z$.  Intuitively, the
estimate~\eqref{eq:5} says that, in the limit of extremely strong (and
unlikely) waves, their probability is dominated by their least
unlikely realization, the instanton.

In practice, the constraint $\eta(L,0)\ge z$ can be imposed by adding
a Lagrange multiplier term to Eq.~\eqref{eq:4}, and it is easier to
use this multiplier as control parameter and simply see
\textit{a~posteriori} what value of $z$ it implies. Concretely, we
perform for various values of $\lambda$ the minimization
\begin{equation}
  \label{eq:6}
  \min_{\psi_0} \left(\tfrac12 \|\psi_0\|^2_C - \lambda
  \eta(L,0)\right)\equiv S_L(\lambda)\,,
\end{equation}
over all the possible realizations of $\psi_0$ (without
constraint). The minimizer $\psi_0^\star(\lambda)$ of this
optimization problem gives the following parametric representation of
$I_L(z)$ versus $z$:
\begin{equation}
  \label{eq:7}
\begin{aligned}
  I_L (z(\lambda)) &= \tfrac12 \|
  \psi^\star_0(\lambda)\|^2_C\,, \\ z(\lambda)
  & = \eta(L,0) = |\psi(L,0)|
  \left(1+\tfrac12 k_0 |\psi(L,0)| \right) \,,
\end{aligned}
\end{equation}
where the last equivalence uses the second order of the Stokes'
series~\eqref{eq:Stokes} at $\theta=0$. It is easy to see from
Eqs.~\eqref{eq:4} and~\eqref{eq:6} that $S_L(\lambda)$ is the Legendre
transform of $I_L(z)$ since:
 \begin{equation} 
   \label{eq:8} 
   \begin{aligned}
     S_L(\lambda) & = \sup_{z\in \RR}(\lambda z - I_L(z))\\
     & = \sup_{z\in\RR}(\lambda z - \tfrac12
     \inf_{\psi_0\in\Lambda(z)}\| \psi_0\|^2_C).
   \end{aligned}
\end{equation}
It is clear from equation~(\ref{eq:5}) that the stochastic sampling
problem is replaced by a deterministic optimization problem, which we
solve numerically as explained next. The trajectory initiated from the
minimizer $\psi_0^*$ of the action will be referred to as the
\textit{instanton trajectory}, and in the following we compare it to
trajectories obtained from the experiment.

\subsection{Numerical aspects}
\label{sec:instanton-num}

In practice, we perform the minimization~\eqref{eq:6} by numerical
gradient descent in the space of the initial condition $\psi_0$, the
gradient being computed by the adjoint formalism. Consequently, for
each iteration of the descent, the NLSE~\eqref{eq:NLSE} needs to be
solved up to $x=L$ for the envelope $\psi$ and its adjoint equation
for the adjoint field $\tilde\psi$. The equation is solved in a time
domain of width $75$ s, much larger than the correlation time of the
wave field (of the order of $10$ seconds), with periodic boundary
conditions in time. The domain is discretized on a lattice of $2^{11}$
equally spaced points. Combined with a cut-off of the initial spectrum
at small amplitude, this leads to $M=89$ modes of the JONSWAP being
relevant for the initial data, as depicted in
Fig.~\ref{fig:jonswap-spectrum}. Eq.~\eqref{eq:NLSE} is numerically
integrated in space by means of a pseudo-spectral exponential
time-differencing method ETDRK2, with a spatial increment of $0.1$
m. More details of the numerical procedure can be found
in~\cite{dematteis2019extreme}.

The minimizer $\psi_0^\star$ of~\eqref{eq:6} identifies the most
likely realization over the distribution of wave shapes at the wave
generator which, evolving deterministically via the NLSE, reaches a
size $\eta(L,0)\ge z$. As saddle point approximation of the
corresponding action, $\psi^\star(z)$ can be considered the {\it
  instanton} of the problem. Here, the large value of $z$ plays the
role of the limiting parameter for the LDP~\eqref{eq:5}. Thus, the
instanton of size $z$ is expected to represent all of the extreme
events $\eta(L,0)\ge z$ to leading order in $z$. Because of this key
property, the instanton is the natural object for the characterization
of the extreme wave events. Note that the knowledge of the instanton
configuration itself can be used as an ingredient for advanced rare
event sampling techniques, such as importance sampling and hybrid
Monte Carlo
approaches~\cite{margazoglou-biferale-grauer-etal:2019}. For the
purpose of this paper, we restrict our analysis to the comparison of
the instanton to the conditioned experimental measurements.

\begin{figure*}
  \centering \includegraphics[width=\linewidth]{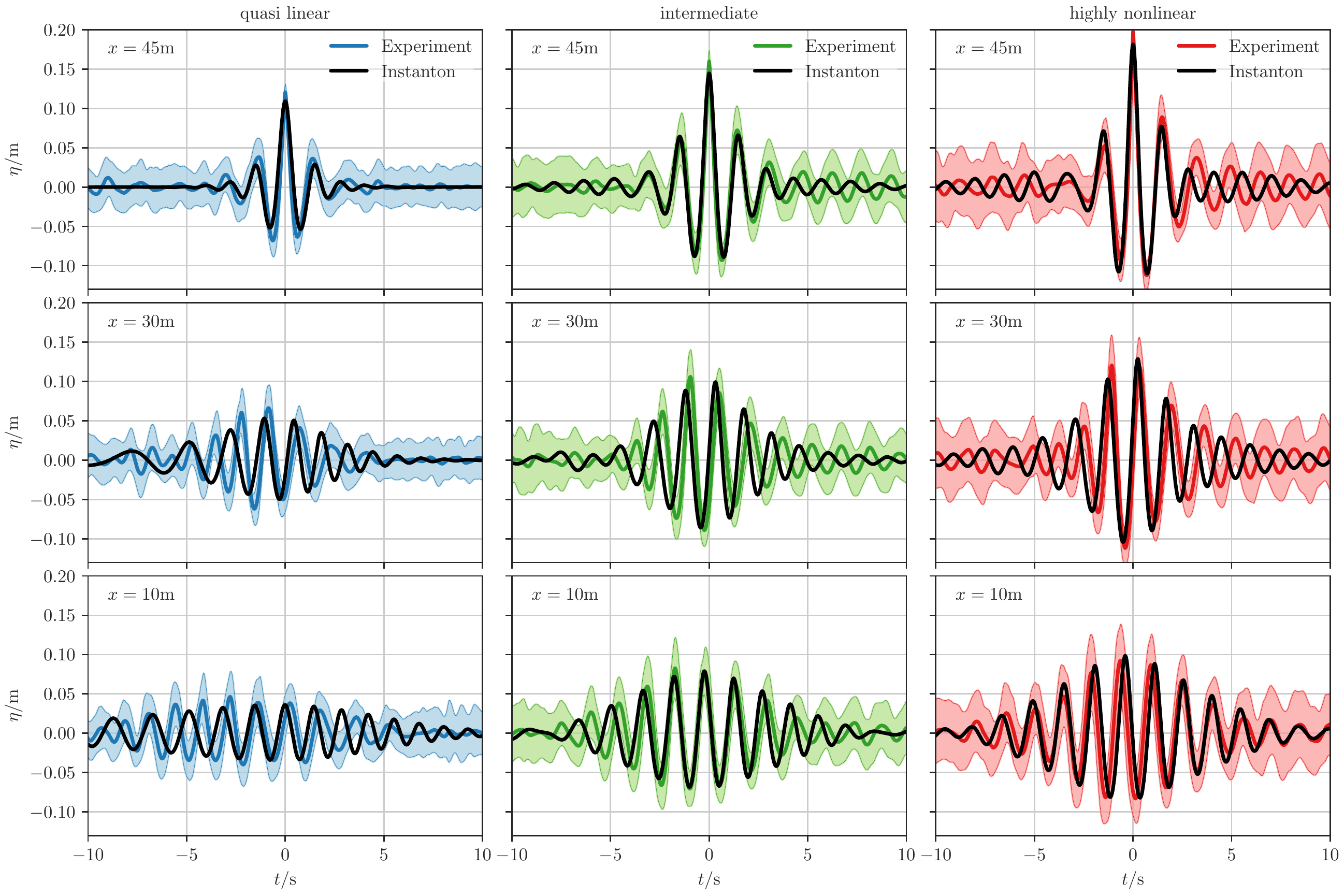}
  \caption{\small Experimental validation of the
    instanton. Snapshots of the instanton during its evolution along
    the channel (black lines) are compared to the mean and standard
    deviation of the experimental rogue wave (color lines), for
    different regimes of nonlinearity. The instanton prediction agrees
    with the experimental mean across all regimes, and captures the
    whole evolution along the channel. This confirms that typical
    rogue waves are well represented by instantons, and the typical
    extreme events collapse onto this most likely one with only small
    fluctuations around them. \label{fig:2}}
\end{figure*}
\begin{figure}
  \centering \includegraphics[width=\columnwidth]{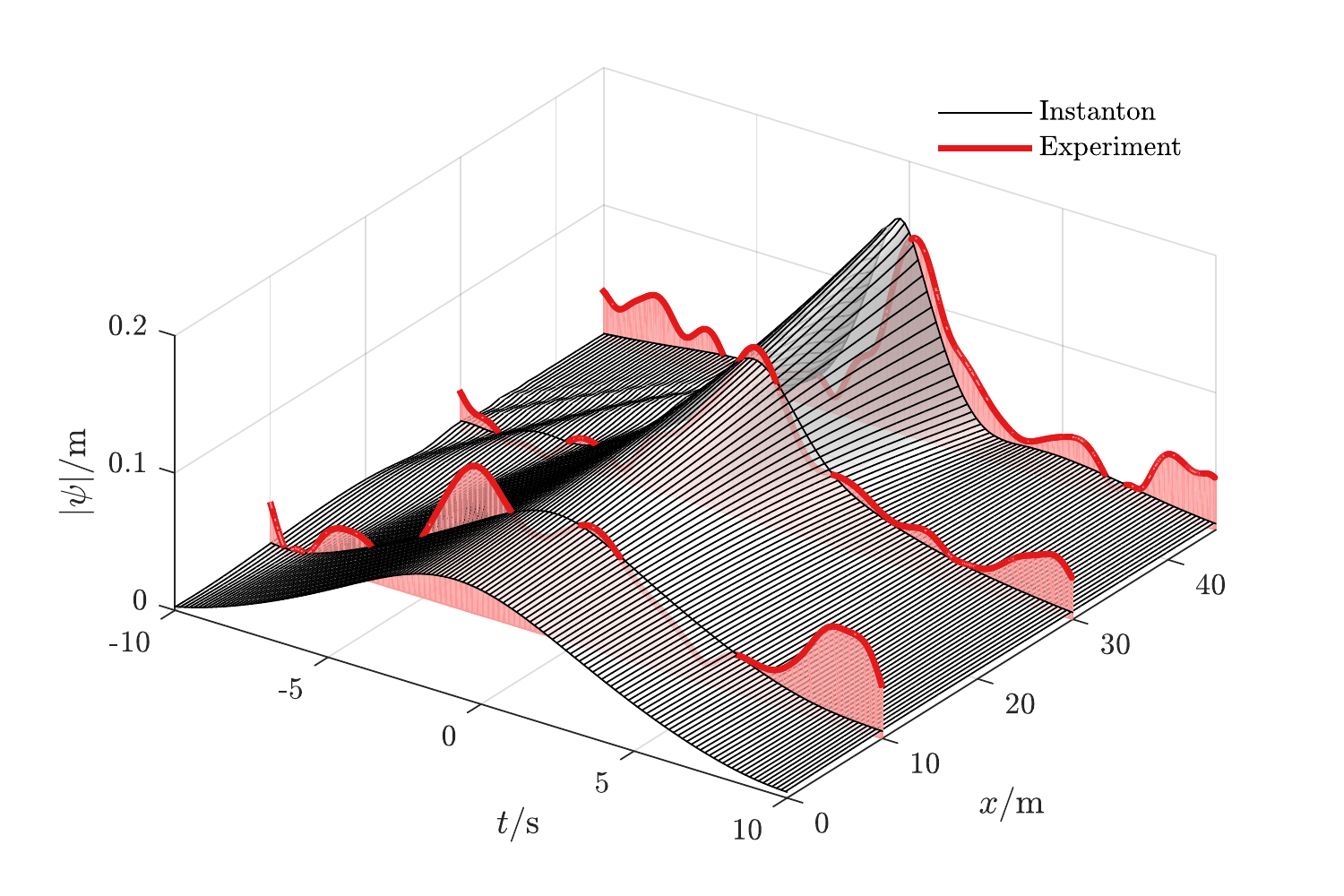}%
  \caption{\small Agreement of the instanton with individual extreme
    events. The evolution of a single realization of an extreme wave
    (red lines) is reasonably approximated by the instanton
    evolution (black/white surface), here for a sample of the highly
    nonlinear data set. In order to capture the focusing pattern in an
    essential way, the envelope $|\psi|$ is plotted instead of the
    surface elevation $\eta$ to remove carrier-frequency
    oscillations.\label{fig:3}}
\end{figure}

\section{Validation of the instanton description}
\label{sec:valid-inst-descr}

In Fig.~\ref{fig:2} we compare the evolution of rogue waves observed
in the experiment and averaged over many realizations to that of the
instanton, both constrained at $x=45$ m. In all cases the instanton
tracks the dynamics of the averaged wave very closely during the whole
evolution. Moreover, in the focusing region the standard deviation
around the mean is small, especially toward the end of the
evolution. This observation in itself is a statement that indeed all
of the rogue waves such that $\eta(L,0)\ge z$ resemble the instanton
plus small random fluctuations. The instanton approximation shows
excellent agreement not only across different degrees of nonlinearity
(and therefore substantially different physical mechanisms), but also
captures the behavior of precursors earlier along the channel.

In Fig.~\ref{fig:3} the envelope evolution of a single realization of
a rogue wave is compared to the instanton evolution at multiple
locations, in the highly-nonlinear case. In the focusing region the
experimental sample shares with the instanton the same overall
structure, needed to allow it to reach an extreme size.

It is worth stressing that the instanton approach captures both the
linear and the fully nonlinear cases, unlike previous theories that
could describe each of these regimes individually but not both. To
make that point, in the next two sections we compare the predictions of
our approach to those of the quasi-determinism and semi-classical
theories that hold in the dispersive and nonlinear regimes,
respectively.

\subsection{Comparison to linear theory}
\label{sec:comp-lin-theory}

\begin{figure*}
  \centering \includegraphics[width=\linewidth]{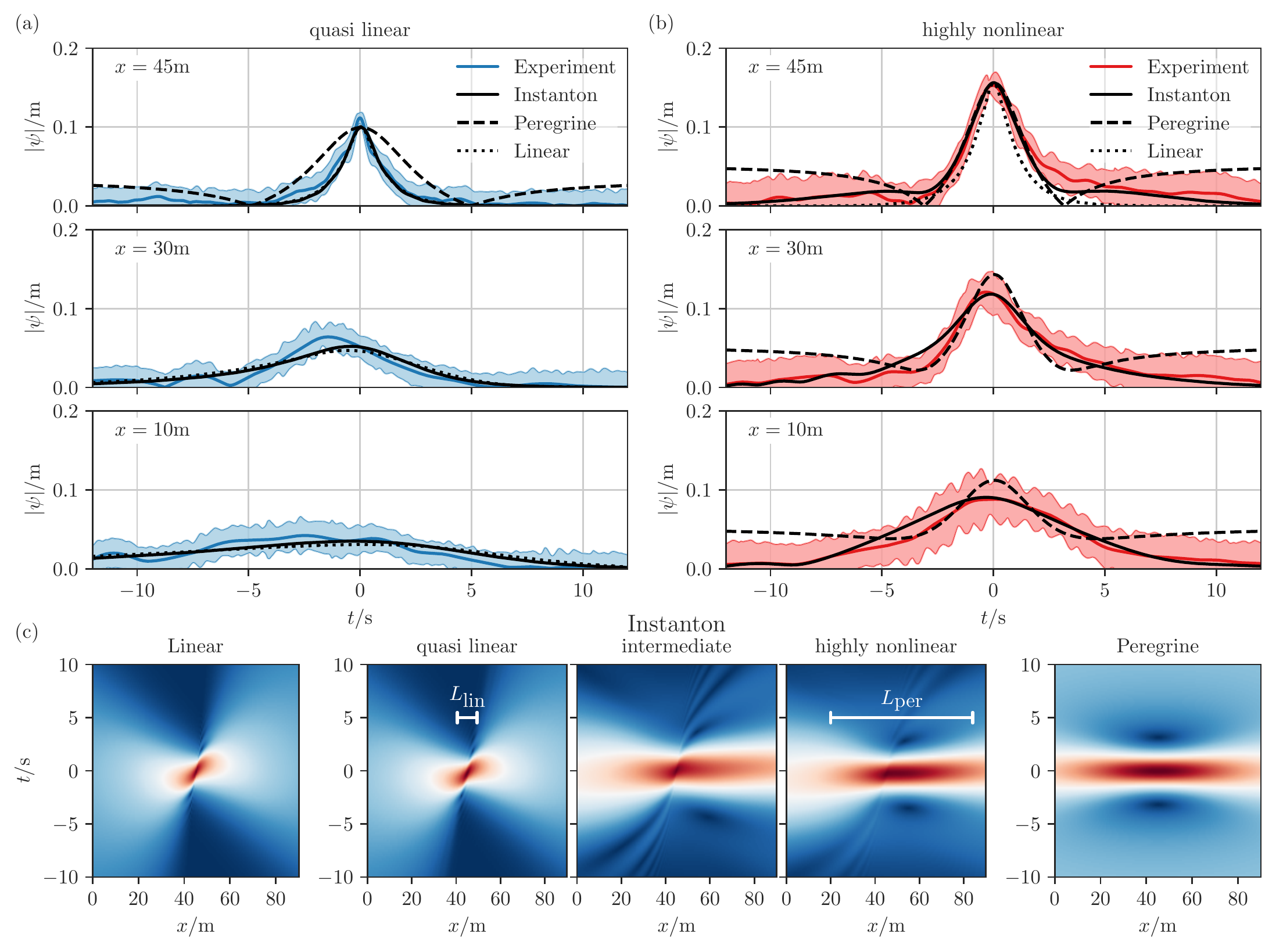}%
  \caption{\small Comparison of the instanton to the predictions of
    the theory of quasideterminism and the semiclassical theory: (a)
    The quasi-linear instanton converges to the linear prediction,
    correctly reproducing the rogue waves averaged over the
    experiments. (b) The highly nonlinear instanton evolution closely
    follows the averaged rogue wave and converges locally to a
    Peregrine soliton around its space-time maximum, as predicted by
    the semi-classical theory, and reproduced by the instanton. The
    linear prediction instead fails, especially around the
    maximum. (c) The contour plots show agreement with the two
    limiting theories and recover the respective dominant length
    scales. In the linear limit, dominated by dispersion, the rogue
    waves arise and decay very rapidly. On the contrary, in the
    semi-classical limit, where nonlinear effects are prevalent, the
    Peregrine-like structure of the extreme event is persistent, with
    a very slow decay. The rogue waves in intermediate regimes display
    both linear and nonlinear features, as shown in the central
    panel.\label{fig:4}}
\end{figure*}

In the linear case, i.e. when the field $\psi(x,t)$ is Gaussian and
stationary, the shape of an envelope time series with a large local
maximum in $t=0$ is expected to be given by the covariance of the wave
field, i.e.~the inverse Fourier transform of the spectrum. This is a
well established result in probability~\cite{lindgren1972local}. In
the oceanographic context, the result was rediscovered in the
$'90$s~\cite{boccotti2000wave} and subsequently tested for some real
quasi-Gaussian wave records in the ocean~\cite{tayfun2007expected},
also accounting for second-order Stokes'
corrections~\cite{fedele2005weakly}. A core result of the theory is
the prediction that conditioning the surface elevation to have a large
maximum, the expected shape of the water surface is given by the
covariance of the wave field, i.e.~the inverse Fourier transform of
the spectrum. The theory is often referred to as the theory of
quasi-determinism, which hereafter we name the linear theory for
simplicity. In our case, such prediction is justified if the nonlinear
focusing effects are small so that the statistics stay close to
Gaussian along the tank, as in the quasi-linear set. Then,
conditioning on a temporal maximum of $\eta(L,0)$ at $x=L$, we can
compute the history of the wave packet by evolving NLSE backward in
space. In Fig.~\ref{fig:4}a this linear prediction is plotted in
comparison with the envelope of the averaged rogue wave for the
quasi-linear set. A good agreement is observed at all spatial points
considered. Moreover, the theoretical instanton found through the
optimization procedure reduces perfectly to the linear prediction,
proving that such result is included in the instanton theory and
represents its limiting linear case.

\subsection{Nonlinear regime and Peregrine solitons}
\label{sec:nonl-regime-peregr}

At the opposite end, in the nonlinear regime, it was recently
shown~\cite{bertola2013universality} that in the zero-dispersion
(semi-classical) regime of the NLSE any single localized pulse on a
vanishing background leads locally to the emergence of a Peregrine
soliton. By scale invariance of the NLSE, such a regime can be
attained whenever an initial condition is characterized by large
enough wave groups for which the nonlinear term dominates over the
dispersive one. In fiber
optics~\cite{suret2016single,tikan2019effect}, emerging Peregrine-like
structures have been observed out of a random background. For the
highly nonlinear case, in Fig.~\ref{fig:4}b we compare the instanton
and the Peregrine soliton reaching the same maximal height $z$ at
$x=45$~m, finding that in the focusing region the two converge to the
same shape, which is also closely followed by the envelope of the
experimental averaged rogue wave. Looking at the event precursor at
earlier $x$, instead, we notice that the experimental mean wave stays
close to the instanton, while it gradually deviates from the Peregrine
soliton. Thus, it appears that the instanton captures the mechanism
underlying the rogue wave events also when nonlinearity rules over
dispersion, tending locally to the Peregrine soliton around the
maximal focusing point, consistently with the regularization of the
gradient catastrophe~\cite{bertola2013universality}.

\subsection{A unified picture of rogue waves}
\label{sec:unif-pict-rogue}

A useful quantification of the effective mechanisms of rogue wave
creation can be obtained by looking at the length scales at play. The
linear length of dispersion is given by
$L_{\text{lin}}=\omega_0^2/(k_0\Omega^2)$, while the characteristic
length associated with the Peregrine soliton is
$L_{\text{Per}}=\sqrt{L_{\text{lin}}L_{\text{nl}}}$
\cite{el2018spontaneous}, where $L_{\text{nl}}=8/(k_0^3H_s^2)$ is the
nonlinear length of modulational instability. These length scales are
clearly visible in space-time contours of the amplitude shown in
Fig.~\ref{fig:4}c, t. In the linear and quasi-linear regimes, the wave
packet has a characteristic length around $L_{\text{lin}}\simeq
9$~m. Thus, we can state that linear superposition dominates and the
expected mechanism leading to the extreme event is the linear
dispersion of a coherent wave packet. The quasi-linear instanton
evolution is almost indistinguishable from the linear
approximation. On the other hand, the extent of the structures in the
highly-nonlinear case agrees with the length
$L_{\text{Per}}\simeq65$~m. The dynamics of the highly nonlinear
instanton clearly converges to the Peregrine dynamics near the
space-time point of maximal focusing, and reproduces the
characteristic isolated ``dips'' of the amplitude observed around the
extreme event. Fig.~\ref{fig:4}c highlights the sharp difference
between the rapidly evanescent linear rogue waves and the more
persistent nonlinear ones. Quite strikingly, the instanton is able to
interpolate between those two limiting regimes, as evidenced by the
intermediate instanton in Fig.~\ref{fig:4}c, which displays features
of both the linear theory and the Peregrine soliton.  Summarizing, the
instanton predicts the shape of rogue waves experimentally observed in
the tank across all parameter regimes.

\subsection{Probability estimates from LDT}
\label{sec:prob-est}

\begin{figure}
\includegraphics[width=\columnwidth]{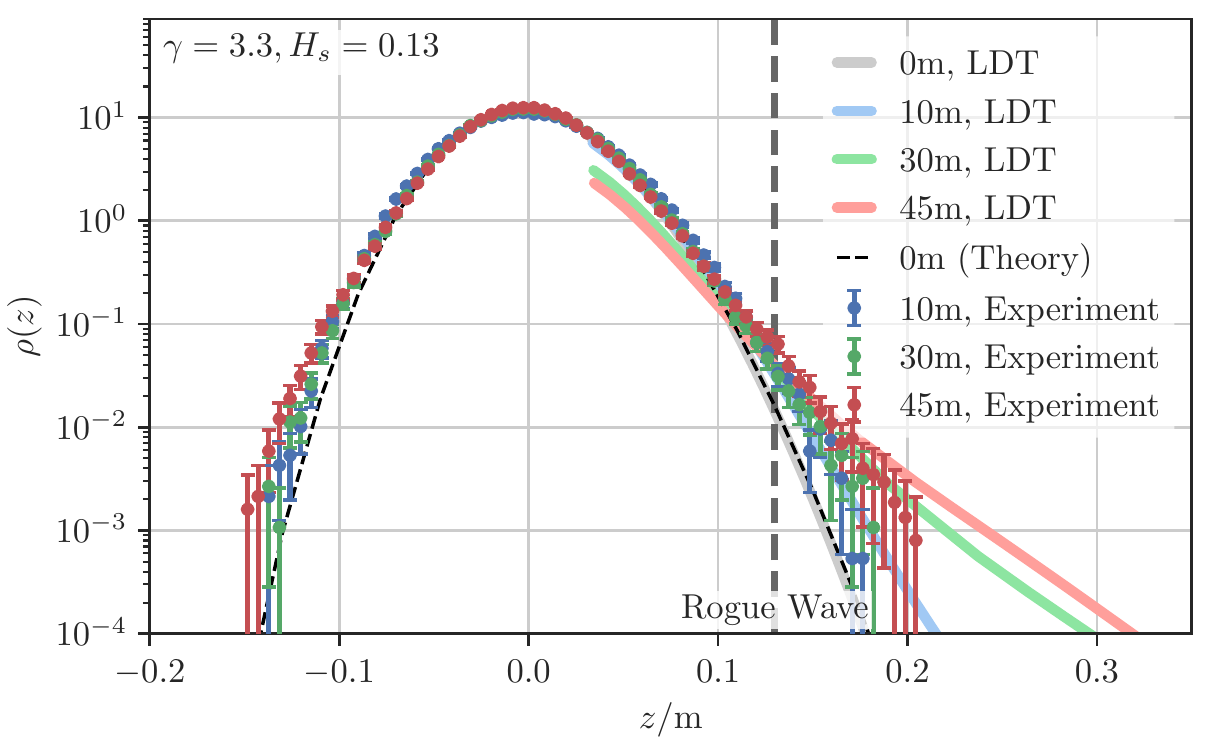}
\caption{\small {
    Comparison of the PDF of the surface elevation $\rho(z) = -P'(z)$
    obtained by binning the data from the experiments and the LDT
    estimates from Eq.~\eqref{eq:5}, showing good agreement in the
    rogue-wave regime (right-hand side of the vertical dashed line,
    indicating the conventional Rogue wave threshold $z=H_s$). The
    figure refers to the intermediate regime. The blue, green and red
    colors indicate data collected at the probes $10$, $30$ and $45$ m
    away from the wave maker, respectively. } \label{fig:5}}
\end{figure}

The analysis so far has addressed the mechanism of rogue-wave
formation, and compared the most likely evolution into an extreme
wave, as predicted by the instanton, to the observed events measured
in the experiment. Since the instanton formalism is based on
probability theory and large deviations, it also allows us to deduce
the tail scaling of the extreme event probability itself
via~\eqref{eq:5}. Indeed, it was shown in \cite{dematteis2018rogue}
that the LDT prediction for the tail of the PDFs match very well those
obtained by brute-force Monte Carlo simulations using NLSE. In the
context of actual experiments, the situation is more
complicated. Despite the large amount of data collected in the
experiments, the far tail of the PDF of the surface elevation is
characterized by a natural cut-off related to the phenomenon of wave
breaking, visually observed during the experiments in the non linear
regimes. The NLSE itself misses such effect that lowers the
probability to observe rogue wave in experiments, especially in the
highly nonlinear regime. As a result, the predictions we can make
about the PDFs of rogue waves are less accurate than those about their
shape. In Fig.~\ref{fig:5} we plot the LDT predictions for the PDF of
the surface elevation, $\rho_L(z) = -P_L'(z)$, in the intermediate
regime at three spatial points with $L=10$, $30$ and $45$~m away from
the wave maker, and compare them with the experimental ones. While the
agreement is reasonable past the height threshold for rogue waves, and
confirms the expected nonlinear tail
fattening~\cite{onorato2005modulational,onorato:2006}, it is difficult
to quantify how accurate these results are because of the problems
mentioned earlier.

\section{Conclusions}
\label{sec:conclusions}

Starting with the pioneering works
in~\cite{dysthe99,Henderson:99,osborne00}, it has been recognized that
nonlinear focusing effects may play an important role in the formation
of rogue waves. Since then, exact solutions of the NLSE, like for
example the Peregrine solution, have been reproduced in controlled lab
experiments \cite{chabchoub2011rogue,kibler2010peregrine} and by now
are considered as prototypes of rogue waves. In random wave fields,
however, our understanding of the development of rogue waves remains
more limited. In strongly nonlinear conditions (semiclassical limit),
assuming a one dimensional propagation described by the NLSE, it has
been shown~\cite{tikan2017universality} that a localized initial
condition leads to the development of extreme waves that can be
locally fitted to the Peregrine solution of the NLSE. While this fit
may suggest the internal mechanism leading to rogue waves in
long-crested, narrow-banded deep seas (neglecting other effects such
as bathymetry, interactions with sea currents, multimodality, etc.,
which may also play a significant role in particular situations) it
says nothing about their likelihood. Such information is instead
intrinsically contained within the instanton framework, allowing for
estimates such as in Fig.~\eqref{fig:5}. To what extent these
nonlinear effects are at work in real directional sea states is also a
difficult question~\cite{fedele2016real, onorato:2009,
  benetazzo2017shape}, in part because of the uncertainty in the
measurements of the directional wave spectrum, especially close to its
peak. If the sea state conditions are not prone for the development of
such nonlinear waves, linear dispersion may still be the dominant one
for generating rogue waves \cite{fedele2016real}. This idea is at the
core of the theory of quasi-determinism (also known as NewWave theory)
that was developed in the early seventies to describe rogue waves in
this linear regime~\cite{lindgren1972local,boccotti2000wave}; it
allows one to determine the shape of the most extreme wave and relate
it to the autocorrelation function.  The two, apparently incompatible,
mechanisms of formation of rogue waves, i.e.~the nonlinear focusing
and the linear superposition, have led to many debates among different
groups of research.

Here we have proposed a unifying framework based on Large Deviation
Theory and Instanton Calculus that is capable to describe with the
same accuracy the shape of rogue waves that result either from a
linear superposition or a nonlinear focusing mechanism. In the limit
of large nonlinearity, the instantons closely resemble the Peregrine
soliton used
e.g. in~\cite{bertola2013universality,tikan2017universality} to
describe extreme events, but with the added bonus that our framework
predicts their likelihood; in the limit of linear waves, the instanton
reduces to the autocorrelation function as obtained in
\cite{lindgren1972local,boccotti2000wave}. A smooth transition between
the two limiting regimes is also observed, and these predictions are
fully supported by experiments performed in a large wave tank with
different degrees of nonlinearity.  These results were obtained for
one dimensional propagation, but there are no obstacles to apply the
approach to two horizontal dimensions, which may finally explain the
origin and shape of rogue waves in different setups, including the
ocean.

\section{Acknowledgments}
M. O. has been funded by Progetto di Ricerca d'Ateneo
CSTO160004. M.O. and G.D. were supported by the ``Departments of
Excellence 2018-2022'' Grant awarded by the Italian Ministry of
Education, University and Research (MIUR) (L.232/2016). E.V.E. was
supported by National Science Foundation (NSF) Materials Research
Science and Engineering Center Program Award DMR-1420073; and by NSF
Award DMS-1522767. M.O. and  E.V.E. were supported by Simons Collaboration
on Wave Turbulence, Award 617006.

\begin{appendix}
\section{Derivation of Eq.~\eqref{eq:10}}
\label{sec:app}

Let
\begin{equation}
  \label{eq:11b}
  \eta_0(t) = \tfrac12 \left(\psi_0(t) e^{-i\omega_0 t} + \bar \psi_0(t)
    e^{i\omega_0 t} \right),
\end{equation}
then, using~\eqref{eq:9}, this can also be written as
\begin{equation}
  \label{eq:14}
  \eta_0(t) = \tfrac12 \int_{-\infty}^\infty \left(\hat \psi_0(\omega)
    e^{i(\omega-\omega_0 ) t} + \Bar{\hat \psi}_0(\omega)
    e^{-i(\omega-\omega_0 ) t} \right) d\omega.
\end{equation}
This implies, using~\eqref{eq:10}, that
\begin{equation}
  \label{eq:15}
  \begin{aligned}
    \langle \eta(t) \eta(t')\rangle & = \tfrac12\int_{-\infty}^\infty
    C(\omega-\omega_0)
    \cos((\omega-\omega_0 ) (t-t')) d\omega\\
    & = \int_{0}^\infty C(\omega) \cos(\omega
      (t-t')) d\omega
  \end{aligned}
\end{equation}
which is consistent with~\eqref{eq:3}.

\end{appendix}

%\bibliographystyle{apsrev4-1-prx}
% \bibliography{bib}
%

\end{document}